\RequirePackage{fix-cm}
\documentclass[twocolumn]{svjour3}          % twocolumn
\smartqed  % flush right qed marks, e.g. at end of proof
\usepackage{graphicx}
\usepackage[misc,geometry]{ifsym}  % Envelope/Letter
\usepackage[numbers]{natbib}			% Numbered References
\usepackage[hyphens]{url}
\usepackage{hyperref}
\usepackage{tikz}
\usepackage{circuitikz}
\usepackage{booktabs}
\usepackage{siunitx}
\usepackage{paralist}
\usepackage{balance}
\usetikzlibrary{backgrounds,intersections,patterns,shapes,arrows}

\hypersetup{
    colorlinks=true,
    linkcolor=blue,
    citecolor=blue,
    urlcolor=blue
}

\journalname{Datenbank-Spektrum}
\begin{document}

\title{On the Diversity of Memory and Storage Technologies}

\author{
        	Ismail Oukid					\and
        	Lucas Lersch
         	%etc.
}

\institute{
              I. Oukid (\Letter) 
              SAP SE, Germany\\
              \email{ismail.oukid@sap.com} \\
              L. Lersch  (\Letter)
              TU Dresden \& SAP SE, Germany\\
              \email{lucas.lersch@sap.com}
}

\date{Published online: 4 June 2018}
  
\maketitle

\begin{abstract}
The last decade has seen tremendous developments in memory and storage technologies, starting with Flash Memory and continuing with the upcoming Storage-Class Memories.
Combined with an explosion of data processing, data analytics, and machine learning, this led to a segmentation of the memory and storage market. Consequently, the traditional storage hierarchy, as we know it today, might be replaced by a multitude of storage hierarchies, with potentially different depths, each tailored for specific workloads.
In this context, we explore in this ``Kurz Erkl\"art'' the state of memory technologies and reflect on their future use with a focus on data management systems.

\keywords{Storage \and Main Memory \and Flash \and SSD \and DRAM \and Storage-Class Memory \and Non-Volatile Memory}

\end{abstract}

%===========================
% ==== Introduction
%===========================
\section{Introduction}
\label{sec:introduction}
\begin{figure*}
\center
\includegraphics[width=\textwidth]{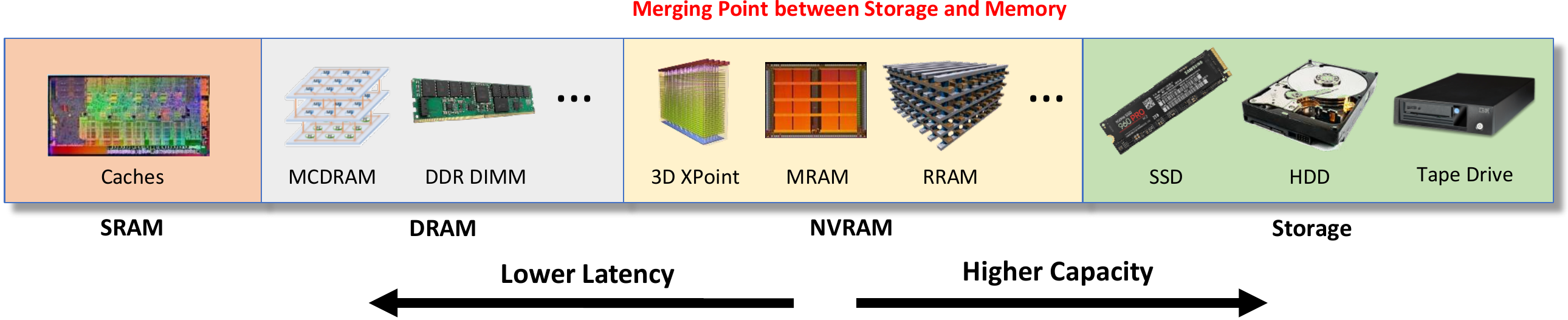}
\caption{\label{fig:memory} Illustration of the diversity of memory and storage technologies (adapted from~\protect\cite{Lehner2017}).}
\end{figure*}

\noindent
The traditional storage hierarchy comprises several layers of memory technologies, ordered from the fastest and least dense to the slowest and densest: CPU caches (SRAM), main memory (DRAM), secondary memory (HDD), and potentially tertiary memory (Tape Drive).
The rise of Flash memory, manufactured in the Solid-State Drive (SSD) form, has pushed HDDs another level down the storage hierarchy: SSDs have successfully superseded HDDs.
However, the rise of novel memory technologies, such as Storage-Class Memories
(SCM), and substantial hardware and software advances in existing ones, such as
Open-Channel SSDs~\cite{open_channel_ssds}, force us to reconsider how we conceive
storage hierarchies.
Indeed, storage system designers are faced with an unprecedented diversity of memory and storage technologies, as illustrated in Figure~\ref{fig:memory}.

The performance of data-intensive applications is directly dependent on the performance of the underlying storage system.
Depending on their host device (servers, smartphones, embedded devices, etc.), and depending on the nature of their data access patterns, these applications may be bound by memory bandwidth, by memory latency, or by energy consumption.
In a real world scenario, a data-intensive application will be constrained by a combination of the three.
This disparity in bandwidth, latency, and energy consumption constraints led hardware manufacturers to segment the market of memory and storage devices into several products, each of which exhibits a different instantiation of the aforementioned three-way trade-off (see more details in Sections~\ref{sec:ram} and~\ref{sec:storage}).

On the one hand, the segmentation and diversification of memory technologies bring the opportunity to build storage systems that are optimized for specific workloads.
On the other hand, building such systems is complex and requires exposing traditionally hardware-managed parts of storage to the application layer (e.g., Open-Channel SSDs).
In the cloud, this task is even more complex as memory and storage resources may be shared by multiple entities, thereby making quality of service a challenging task for cloud providers.

The first ``Kurz Erkl\"art'' of this series explored the diversity of computing units and the opportunities they bring to data management systems~\cite{Lehner2018}. In this second ``Kurz Erkl\"art'', we give an overview of the state of advancement --
from a systems developer point of view -- of memory and storage technologies
and their impact on data management systems.
To do so, we organize the remainder of this paper as follows: Section~\ref{sec:ram} and Section~\ref{sec:storage} discuss recent developments in main memory and storage technologies, respectively.
Thereafter, Section~\ref{sec:scm} introduces Storage-Class Memory, highlights its opportunities, and underlines its challenges.
Finally, Section~\ref{sec:summary} summarizes the paper and outlines future breakthroughs lying ahead of us.

%===========================
% ==== Random Access Memory
%===========================
\section{Random Access Memory}
\label{sec:ram}
\noindent
There are two main types of Random Access Memory (RAM): Static RAM (SRAM) and Dynamic RAM (DRAM). SRAM requires six transistors per memory cell and relies on changing the direction of the current to read and write memory cells. In contrast, DRAM requires only one transistor and one capacitor which is used to hold the charges. Therefore, DRAM is much simpler, denser, and cheaper (since it requires six times less transistors) than SRAM. However, since DRAM's capacitors produce current leakage, its memory cells must be constantly refreshed -- hence the name ``Dynamic'' RAM. While DRAM is denser and simpler to produce, SRAM offers a much lower access latency and a much higher bandwidth.  Therefore, SRAM is usually used for the smaller CPU caches whose performance is critical, while DRAM is used for the larger main memory. Since SRAM is embedded on-chip and inflexible, we focus in the remainder of this section on DRAM.

The DRAM market is currently dominated by Samsung, Micron, and SK Hynix; they own more than 95\% of the market share~\cite{dram_market}. Furthermore, the market is segmented into many categories, each of which is tailored for specific application domains:
\begin{compactitem}
\item \textbf{Double Data Rate (DDR) DRAM} is targeted at Complex Instruction Set Computers (CISC), which can issue multiple instructions in a single cycle, such as CPUs found in desktops and servers. Therefore, it is optimized to handle parallel, small-sized memory requests using a typically 64-bit memory bus.
\item \textbf{Low Power DDR (LPDDR)} offers very low power consumption and is targeted at smaller devices such as smartphones, tablets, and laptops.
\item \textbf{Graphics DDR (GDDR)} is optimized for GPU workloads, or more generally, for Reduced Instruction Set Computers (RISC) that issue a single instruction per cycle. It differs significantly from DDR in that it has a wider memory bus (up to 256-bit wide) which allows it to provide much higher bandwidth. However, it does not handle well parallel non-adjacent memory requests.
\item \textbf{High Bandwidth Memory (HBM)} is a variant of DRAM that provides much higher bandwidth than GDDR thanks to its 3D design: multiple layers of DRAM are stacked together and accessed through a very wide memory bus (typically 1024-bit wide). It is mainly targeted at high-end GPUs and servers.
\item \textbf{Multi-Channel DRAM (MCDRAM)} is a type of HBM introduced by Intel in its second generation Xeon Phi processors\footnote{Intel has discontinued its Xeon Phi series, albeit some of its concepts have converged with the Xeon Scalable series.}~\cite{xeon_phi}. It is a high-bandwidth, low-capacity DRAM that can be used as a software-managed fast buffer between CPU caches and main memory to accelerate analytical workloads.
\item \textbf{Hybrid Memory Cube (HMC)} is another promising high-bandwidth, low-capacity 3D-stacked DRAM and a competitor of MCDRAM. Its application domains include high-end computing and networking.
\end{compactitem}
Each category, with the exception of MCDRAM and HMC, has improved over multiple generations, the latest being DDR5, GDDR6, LPDDR4X, and HBM2.

DDR DRAM is by far the category that offers the highest capacity.
It is also the most relevant for database systems. While the cost per bit of DDR DRAM has steadily decreased over the years, the capacity and bandwidth per core have worsened~\cite{Mutlu2013}.
As a matter of fact, it is intrinsically hard to further increase the density of DRAM~\cite{ITRS}: The smaller the DRAM cell, the more it leaks energy which interferes with the state of neighboring cells, thereby exponentially increasing error rates.
Another concern is that a significant share of data-centers energy consumption is attributed to DRAM~\cite{Dayarathna2016}, either directly or indirectly (e.g., through the cooling system).
Consequently, DRAM no longer satisfies the demand for ever-increasing main-memory capacities.

%===========================
% ==== Non-Volatile Storage
%===========================
\section{Non-Volatile Storage}
\label{sec:storage}

Flash-based solid-state drives (SSDs) were introduced in the early 90's and,
while initially there were many challenges to be addressed, nowadays they have
proven to be a usable and reliable technology.
Similar to hard disk drives (HDDs), SSDs are non-volatile block-based devices. In contrast to HDDs, SSDs are purely electronic storage devices, i.e., without any moving parts, and with a much better performance.
The purely electronic nature of SSDs enables a higher degree of parallelism, which also reduces the performance gap between sequential and random accesses, an important aspect that had to be considered when developing systems for HDDs.
Furthermore, modern SSDs present an IO latency of tens of microseconds, while HDDs still have access latencies in the order of milliseconds.

\begin{figure}
	\center
	\includegraphics[width=\columnwidth]{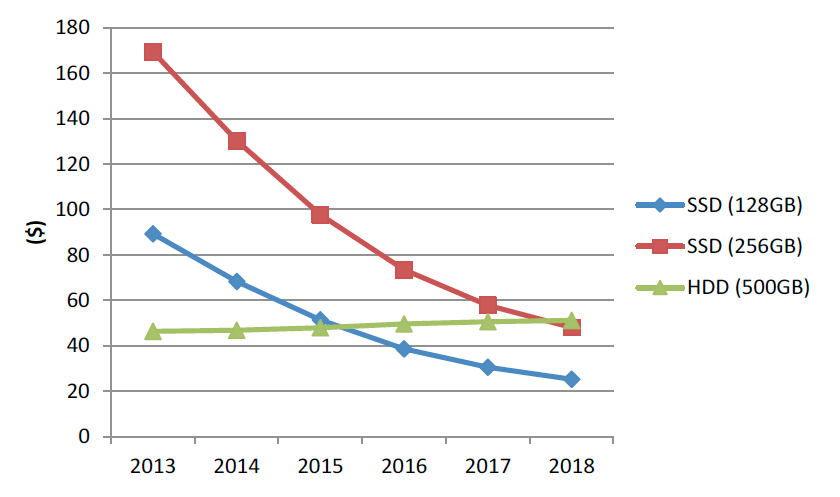}
	\caption{\label{fig:prices} Worldwide SSD and HDD average selling price \cite{sandisk:prices}.}
\end{figure}

Initial disadvantage of SSDs were higher costs and limited write endurance.
However, today SSDs offer enough write endurance and on-chip wear-leveling
that most application do not have to worry about such issues.
A study~\cite{blog:wearout} with 12 different SSDs showed that most of them only reach a wear-out scenario after about five years of regular usage (40 GB writes per day).
The five years interval is similar to the warranty time offered by most HDD manufacturers.
Figure~\ref{fig:prices} shows the average selling prices of SSDs and HDDs.
While SSDs are still more expensive, the price is dropping significantly, while the price of HDDs has stabilized in the past years.

Another aspect worth noting is the power consumption. For consumer-level storage devices, SSDs tend to consume much less power than HDDs. However, this is not necessarily the case for high-capacity, enterprise-level storage devices. For example, according to their respective specifications, the Seagate Exos X12 HDD~\cite{seagate:exosx12} consumes 9.3 Watts during operation, while the Intel DC P4510 SSD~\cite{intel:p4510} consumes 16 Watts. Nevertheless, the power consumption relative to the offered performance makes SSDs a much more attractive option in terms of power savings. Finally, although different flash technologies exist, NAND flash memory became the standard in modern SSDs due to its higher density and better endurance, which translates into lower costs.

\subsection{Density}
Although most modern SSDs are based on NAND flash memory, they differ on the
way these cells are organized internally.
While the early designs stored a single bit per cell in a Single-Level Cell structure (SLC), later improvements were made to increase the density with Multi-Level Cell (MLC) and Triple-Level Cell (TLC), storing respectively two and three bits per cell.
The increased density of MLC and TLC comes at the cost of reduced write performance, higher power consumption, and lower cell endurance.
Most modern enterprise SSDs are either MLC or TLC, as the cost of SLC is prohibitive for systems handling large amounts of data.

Even with the higher densities enabled by MLC and TLC, it became harder to
further scale capacity, since the device becomes much more error prone
as more bits are packed in a single cell and fewer electrons trapped in the
cell correspond to a bit.
Manufacturers have solved this issue by stacking cells vertically, enabling more cells while maintaining the same surface on a single die.
This technology is known as 3D NAND Flash.

In comparison to 2D geometries, the vertical stacking of cells can increase the
capacity of SSDs by two orders of magnitude.
The vast majority of modern SSDs are based on 3D flash memory.
Furthermore, a recent study~\cite{DBLP:conf/fast/SchroederLM16} based on data collected over 6 years in Google data centers showed that the reliability of modern MLC devices is comparable to that of SLC devices, reducing the range of use cases for SLC.
The same study also shows that, while SSDs had a lower replacement rate than HDDs, the rate of uncorrectable errors was higher.

\subsection{Performance}
SSD flash cells are organized into packages of a certain capacity.
While this capacity can be increased by adding more dies on a flash package, this has a negative impact on performance, as the access times of a single die increase.
However, the decrease in performance is compensated by increasing the internal
parallelism within and across packages.
The SSD controller uses multiple individual channels to communicate to the packages.
Thanks to this intrinsic parallelism, SSDs achieve a much higher bandwidth by striping data and interleaving accesses across packages.

The increase in performance made it necessary to adapt both hardware and
software interfaces, as these became the bottleneck.
Initially SSDs adopted the same Serial ATA (SATA) interface found in most HDDs.
However, as the SATA interface could not keep up with the potential bandwidth, many manufacturers started to offer SSDs with a more performant PCI Express (PCIe) interface.
PCIe was later established with the standardization of the NVMe specification.
While SATA SSDs are still sold on the market, the tendency is that these will be replaced by PCIe/NVMe in the near future.
For instance, an Intel DC S4600 Series (SATA)~\cite{intel:s4600} offers a bandwidth of up to 500 MB/s while the Intel DC P4510 (PCIe)~\cite{intel:p4510} offers a bandwidth of up to 3000 MB/s.

Another alternative to SATA offered by some manufacturers is the Serial
Attached SCSI (SAS) interface.
While SAS performance is much better than SATA, but generally worse than PCIe, its key advantage over PCIe is the flexibility regarding extending a server's storage capacity simply by adding additional SAS devices to multi-port arrays. PCIe is not easily extensible: the devices have to be connected directly to the chipset.

While the underlying media (3D NAND flash memory) and hardware interface (PCIe)
are the same for most SSDs, other aspects have critical impact on the
performance characteristics of the final product.
Some examples are the degree of parallelism, amount of cache, and the controller.
The controller itself plays a major role, as it is responsible for multiple functionalities such as error detection and correction, bad block mapping, compression, wear leveling, and garbage collection.
Therefore, it is common for a single manufacturer to offer a wide range of products aimed at specific scenarios (e.g., read/write optimized, mixed workloads, large capacities, and low latency).

Finally, software changes are
also required to fully exploit the potential of modern SSDs. To that aim, efforts exist to allow the application to bypass the SSD controller and have
full control over the behavior of the device. This is achieved by exposing inner SSD interfaces and allowing
the application to optimize aspects such as parallelism and wear-leveling for specific use cases.
This class of devices is known as Open-Channel SSDs and are already supported by Linux kernel through the LightNVM subsystem~\cite{Bjorling2017}.

\subsection{Recent and Future Developments}
As the performance of SSDs keep improving, the bottleneck shifts to other
components in the system's stack.
The vast majority of modern storage devices still operate through a block-level interface.
The mismatch between the application representation (e.g., objects) and the block representation requires data to be converted when reading or writing to the storage device.
To lift the overhead introduced by this conversion, Samsung has announced a Key-Value SSD~\cite{samsung:kvssd} which implements the usual logic present in key-value stores in the SSD firmware, allowing the application to interact with the device in a much simpler way. Moreover, Intel announced its Optane~\cite{intel:p4800x} line of products.
Different than 3D NAND flash, Optane is based in the Intel 3D XPoint memory
technology. The 3D XPoint technology does not only promise to offer 3--10 times lower latency, but also a much higher endurance than 3D NAND Flash.

While the performance of accessing local NVMe SSDs has greatly improved, leveraging these improvements over the network became a challenge.
Recent work~\cite{Klimovic2017} has proposed a system with a tight coupling of network and storage in order to fully leverage NVMe SSDs performance and enable remote access latencies comparable to local ones. Many works such as Li et al.~\cite{li2010tree} and Levandoski et al.~\cite{Levandoski2013} have presented data structures designed to better exploit the characteristics of flash-based SSDs. More specific to the context of database systems, works such as Hardock et al.~\cite{Hardock2017, Hardock2013} have exploited native flash management to reduce write amplification in presence of small updates.

\subsection{HDD \& Tape}
For many decades, HDDs have been the default storage media for
the vast majority of systems.
Rapid advancements in SSD technologies raise the question whether HDDs will become obsolete in the near future. Even with lower performance characteristics, HDDs still offer a lower price per
Gigabyte and potentially higher reliability than SSDs, which would make them a
good alternative for archival and backup storage.
In such a scenario, HDDs would compete directly with tape-based storage, which surprisingly-enough is still around.

A recent study~\cite{ADMS:fiveminute} compared the 
characteristics of HDD and tape for archival and backup purposes.
We highlight here three important observations.
First, density of tape-based storage has been increasing in a higher rate than that of HDDs.
Second, tape-based storage has a much lower idle power consumption. Third, the bandwidth of sequential access on tape can match that of modern HDDs.
Based on these observations, one may wonder whether HDDs will still have a place in the storage hierarchy in the near future, as they are currently not fast enough to compete with SSDs, and not economic enough to compete with tape.

%===========================
% ==== Storage-Class Memory
%===========================
\section{Storage-Class Memory}
\label{sec:scm}

\begin{quote}
\textit{``The arrival of high-speed, non-volatile storage devices, typically referred to as storage-class memories (SCM), is likely the most significant architectural change that datacenter and software designers will face in the foreseeable future.''}~\cite{Nanavati2016}.
\end{quote}

\noindent
Storage-Class Memory\footnote{SCM is also referred to as Persistent Memory, Non-Volatile RAM (NVRAM), or simply Non-Volatile Memory.} (SCM) is a class of novel memory technologies that exhibit characteristics of both storage and main memory: They combine the non-volatility, density, and economic characteristics of storage (e.g., flash) with the byte-addressability and a latency close to that of DRAM (albeit higher). Examples of such memory technologies include Resistive RAM~\cite{rram} (researched by SK Hynix, SanDisk, Crossbar, Nantero), Magnetic RAM~\cite{mram} (researched by IBM, Samsung, Everspin), and Phase-Change Memory~\cite{pcm} (researched by IBM, HGST, Micron/Intel). Table~\ref{tab:latencies} summarizes the projected characteristics of these technologies and compares them to those of SLC Flash and DRAM. In particular, Intel and Micron announced an SCM technology, called 3D XPoint\footnote{Intel and Micron did not disclose so far the technology that 3D XPoint is based on, albeit it has been speculated that it is based on Phase-Change Memory~\cite{blog:pcm}.}~\cite{3d_xpoint}, in the Dual Inline Memory Module (DIMM) form factor. SCM technologies are expected to exhibit asymmetric latencies, with writes being noticeably slower than reads, and limited write endurance (although SCM may be significantly more durable than flash memory, e.g., 3D XPoint by three orders of magnitude). Moreover, SCM will be denser than DRAM, yielding larger memory capacities. Finally, in contrast to DRAM that constantly consumes energy to refresh its state, idle SCM does not consume energy -- only active cells do. Hence, SCM has the potential to lift the scalability issues of DRAM, both in terms of capacity and energy consumption.

\begin{figure}
\center 
\includegraphics[width=0.8\columnwidth]{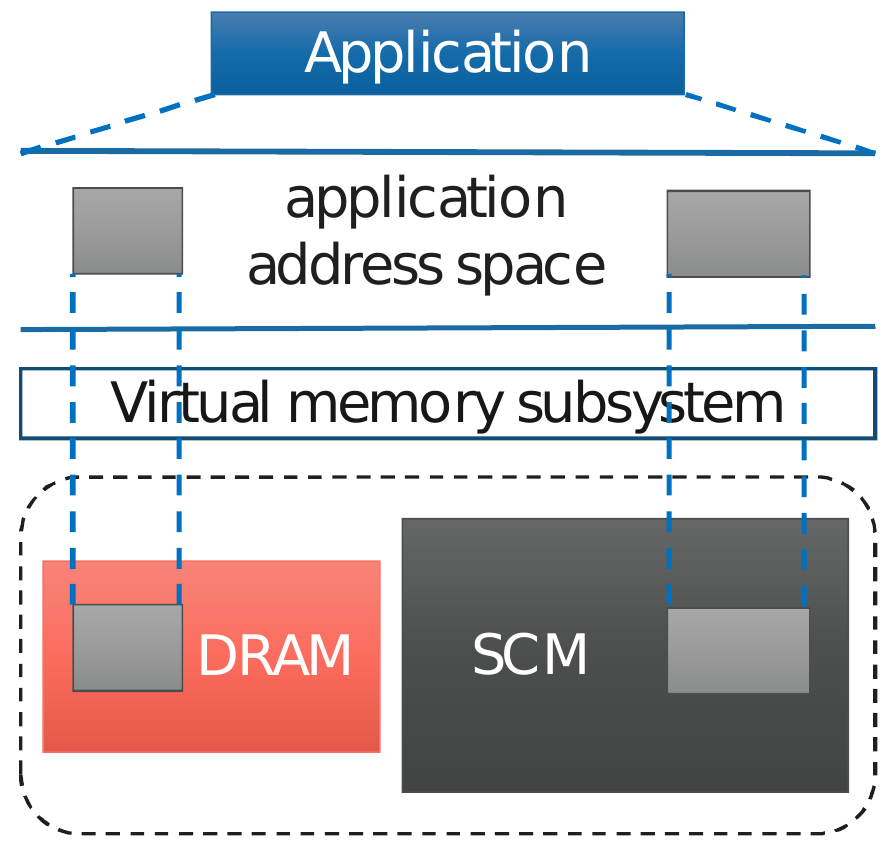}
\caption{\label{chap1:fig:scmArchitected} SCM is mapped directly into the address space of the application, allowing direct access with load/store semantics.}
\end{figure}

\begin{table*}
\center 
  \begin{tabular}{c|lllll}
Parameter & SLC Flash & DRAM & PCM & STT-MRAM & RRAM \\
\midrule
Read~Latency& \SI{25}{\micro\second}& \SI{50}{\nano\second}& \SI{50}{\nano\second}& \SI{10}{\nano\second}& \SI{10}{\nano\second}\\
Write~Latency& \SI{500}{\micro\second}& \SI{50}{\nano\second}& \SI{500}{\nano\second}& \SI{50}{\nano\second}& \SI{50}{\nano\second}\\
Byte-addressable& No& Yes& Yes& Yes& Yes \\
Endurance& 10$^4$--10$^5$& $>$10$^{15}$& 10$^8$--10$^9$& $>$10$^{15}$& 10$^{11}$ \\
Density& High & Low & Medium & Low & High \\
\end{tabular}
\caption{Comparison of the characteristics of SLC Flash and DRAM with those of several SCM candidates~\protect\cite{Mittal2016}.}
\label{tab:latencies}
\end{table*}

Given its unique characteristics, SCM can serve as fast storage or as DRAM replacement.
However, while SCM is projected to be cheaper than DRAM, it will be at first too expensive to replace flash.
Additionally, it will be too slow at first to replace DRAM.
Nevertheless, we foresee that SCM will be invaluable in extending main-memory capacity in large scale-up systems.
Additionally, it can serve as a cheaper DRAM alternative when performance is not paramount.
We argue, however, that these use cases do not harness the full potential of SCM: They do not exploit its non-volatility.
A third option would be to use SCM as \emph{persistent main memory}, i.e., as memory and storage at the same time.
Given its byte-addressability and low latency, processors will be able to access SCM directly with load/store instructions.
Both Microsoft Server~\cite{ms_nvdimm} and Linux~\cite{linux_nvdimm} already support this access method, called Direct Access (DAX), by offering zero-copy memory mapping that bypasses DRAM and grants the application layer direct access to SCM, as illustrated in Figure~\ref{chap1:fig:scmArchitected}.

While SCM brings unprecedented opportunities as a potential universal memory,
it fulfills the \emph{no free lunch} folklore conjecture and raises
unprecedented challenges as well.
To store data, software has traditionally assumed block-addressable devices, managed by a file system and accessed through main memory.
The programmer holds full control over when data is persisted and the file system takes care of handling partial writes, leakage problems, and storage fragmentation.
As a consequence, database developers are used to ordering operations at the logical level, e.g., writing an undo log before updating the database. SCM invalidates these assumptions: It becomes possible to access, read, modify, and persist data in SCM using load and store instructions at a CPU cache-line granularity.
The journey from CPU registers to SCM is long and mostly volatile, including store buffers and CPU caches, leaving the programmer with little control over when data is persisted. Even worse, compilers and CPUs might speculatively reorder writes.
Therefore, there is a need to enforce the order and durability of SCM writes at the system level (in contrast to the logical level) using persistence primitives, such as memory barriers and cache-line flushing instructions, often in a synchronous way.
This, in turn, creates new failure scenarios, such as missing or misplaced persistence primitives, which can lead to data corruption in case of software or power failure.
As a consequence, leveraging SCM as persistent main memory requires devising a novel programming model.

The last few years have seen a surge in research efforts that investigate how database systems can leverage SCM as persistent main memory. These research efforts can be categorized into: SCM memory management~\cite{pmdk, pallocator, Renen2018, Lersch2017b}, SCM-based persistent data structures~\cite{Yang2015b, Chen2015, fptree, Lee2017}, optimizing database algorithms for SCM~\cite{Chen2011, Viglas2014}, new testing frameworks for SCM-based software~\cite{Lantz2014, nvm_testing}, improving the database logging infrastructure~\cite{Fang2011, Wang2014, Huang2014}, and finally exploring novel, SCM-enabled database storage architectures~\cite{sofort2, Arulraj2016, Kimura2015}.

%===========================
% ==== Conclusion
%===========================
\section{Summary and Outlook}
\label{sec:summary}
\noindent
In this ``Kurz Erkl\"art'', we briefly explored the increasing diversity in memory and storage technologies, and highlighted the rise of SCM as a potential universal memory.
While this diversity brings opportunities for building workload-optimal memory and storage systems, it also pushes more complexity from the hardware layer to the software layer by exposing low-level hardware management features that were traditionally transparent to systems developers.
Nevertheless, cloud providers can abstract away this complexity for cloud users through virtualization. Ideally, it should be possible to build custom memory hierarchies simply by provisioning the desired resources in the cloud. Ensuring quality of service will be the biggest challenge to achieve this vision.

In addition to SCM, more groundbreaking innovations await on the horizon.
First, Processing In Memory (PIM), which requires embedding compute logic on memory devices, is becoming an attractive hardware acceleration method.
For instance, Borumand et al.~\cite{Boroumand2018} showed that PIM can halve both energy consumption and execution time by reducing data movement in widely-used Google customer workloads.
Second, two large consortia, Gen-Z~\cite{genz} and OpenCAPI~\cite{opencapi}, proposed novel memory communication protocols that enable a scalable and flexible hardware topology accounting for heterogeneous memories and accelerators. These and other advancements promise to keep research in memory and storage systems an exciting field!

\\[10pt]

\balance
\bibliographystyle{spbasic}      % basic style, author-year citations
\bibliography{bibliography}   % name your BibTeX data base

\end{document}